# Adjusted Logistic Propensity Weighting Methods for Population Inference using Nonprobability Volunteer-Based Epidemiologic Cohorts

Lingxiao Wang[1], Richard Valliant[2], and Yan Li[1*]

[1]The Joint Program in Survey Methodology, University of Maryland, College Park, U.S.A.

[2]Research Professor Emeritus at the University of Michigan and University of Maryland

*Address correspondence to Yan Li, 1218 Lefrak Hall, 7521 Preinkert Dr, College Park, MD 20742; email: yli6@umd.edu

**Abstract (208 out of 250 words)**

Many epidemiologic studies forgo probability sampling and turn to nonprobability volunteer-based samples because of cost, response burden, and invasiveness of biological samples. However, finite population inference is difficult to make from the nonprobability sample due to the lack of population representativeness. Aiming for making inferences at the population level using nonprobability samples, various inverse propensity score weighting (IPSW) methods have been studied with the propensity defined by the participation rate of population units in the nonprobability sample. In this paper, we propose an adjusted logistic propensity weighting (ALP) method to estimate the participation rates for nonprobability sample units. Compared to existing IPSW methods, the proposed ALP method is easy to implement by ready-to-use software while producing approximately unbiased estimators for population quantities regardless of the nonprobability sample rate. The efficiency of the ALP estimator can be further improved by scaling the survey sample weights in propensity estimation. Taylor linearization variance estimators are proposed for ALP estimators of finite population means that account for all sources of variability. The proposed ALP methods are evaluated numerically via simulation studies and empirically using the naïve unweighted National Health and Nutrition Examination Survey III sample, while taking the 1997 National Health Interview Survey as the reference, to estimate the 15-year mortality rates.

## 1. INTRODUCTION

In the big data era, assembling volunteer-based epidemiologic cohorts within integrated healthcare systems that have electronic health records and a large pre-existing base of volunteers are increasingly popular due to their cost-and-time efficiency, such as the UK Biobank in the UK National Health Service. [1] However, samples of volunteer-based cohorts are not randomly selected from the underlying finite target population, and therefore cannot well represent the target population. As a result, the naïve sample estimates obtained from the cohort can be biased for the finite population quantities. For example, the estimated all-cause mortality rate in the UK Biobank was only half that of the UK population, [2] and the Biobank is not representative of the UK population with regard to many sociodemographic, physical, lifestyle and health-related characteristics.

Aiming for making inferences at the population level using nonprobability samples, various propensity-score weighting, and matching methods have been proposed to improve the population representativeness of nonprobability samples, by using probability-based survey samples as external references in survey research. [3-6]

Inverse propensity score weighting (IPSW) methods have been studied with the propensity defined by the participation rate of population units in the nonprobability sample. We review two methods—both assume that the units in the nonprobability sample are observed according to some random, but unknown, mechanism. Because that mechanism is unknown, the inclusion probability of each unit must be estimated. As described in section 2, all methods are based on estimating a population log-likelihood, although the methods differ in their details. Valliant and Dever [7] estimated participation rates by fitting a logistic regression model to the combined nonprobability sample and a reference, probability sample. Sample weights for the probability sample were scaled by a constant so that the scaled probability sample was assumed to represent the complement of the nonprobability sample. Each unit in the nonprobability sample was assigned a weight of one. This results in the sum of the scaled weights in the combined probability plus nonprobability sample being an estimate of the population size. This method will be referred to as the rescaled design weight (RDW) method. The participation rate for each nonprobability sample unit was estimated by the inverse of the estimated inclusion (or participation) probability.

The RDW estimator is biased especially when the participation rate of the nonprobability sample is large, as noted by Chen et al. [4] As a remedy, Chen et al [4] estimated the participation rate by manipulating the log-likelihood estimating equation in a somewhat different way. The resulting estimator, denoted by CLW, is consistent and approximately unbiased regardless of the magnitude of participation rates. Compared to the CLW method, which requires special programming, the RDW method has the advantage of easy implementation by ready-to-use software such as R, Stata, or SAS. Survey practitioners can simply fit a logistic regression model with scaled survey weights in the probability sample to obtain the estimated participation rates.

In this paper, we propose an adjusted logistic propensity weighting (ALP) method to estimate the participation rates for nonprobability sample units. Like the CLW, the proposed ALP method relaxes the assumptions required by the RDW method, [7,8] by formulating the method in an innovative way. As in the RDW method, the proposed ALP method retains the advantage of easy implementation by fitting a propensity model with survey weights in ready-to-use software. Taylor linearization variance estimators are proposed for ALP estimates that account for variability due to differential pseudo-weights in the nonprobability sample, complex survey design of the reference probability survey, as well as the estimation of the propensity scores. Moreover, the proposed ALP method is proved, analytically and numerically, to be as or more efficient than the CLW method and can flexibly scale the probability sample weights for propensity estimation to further improve efficiency.

## 2. METHODS

### 2.1. Basic setting

Let $FP = \{1, \cdots, N\}$ represent the finite population with size $N$. We are interested in estimating the finite population mean $\mu = N^{-1} \sum_{i \in FP} y_i$. Suppose a volunteer-based nonprobability sample $s_c$ of size $n_c$ is selected from $FP$ by a self-selection mechanism, with $\delta_i^{(c)}$ ($= 1$ if $i \in s_c$; 0 otherwise) denoting the indicator of $s_c$ inclusion. The underlying participation rate of nonprobability sample for a finite population unit is defined as

$$\pi_i^{(c)} = P(i \in s_c \mid FP) = E_c\left\{\delta_i^{(c)} \mid y_i, x_i\right\}, \quad i \in FP$$

where the expectation $E_c$ is with respect to the nonprobability sample selection, and $x_i$ is a vector of self-selection variables, i.e., covariates related to the probability of inclusion in $s_c$. The corresponding implicit nonprobability sample weight is $w_i = 1/\pi_i^{(c)}$ for $i \in FP$.

We consider the following assumptions for the nonprobability sample self-selection.

**A1**. The nonprobability sample selection is uncorrelated with the variable of interest given the covariates, i.e., $\pi_i^{(c)} = E_c\left\{\delta_i^{(c)} \mid y_i, x_i\right\} = E_c\left\{\delta_i^{(c)} \mid x_i\right\}$ for $i \in FP$.

**A2**. All finite population units have a positive participation rate, i.e., $\pi_i^{(c)} > 0$ for $i \in FP$.

**A3**. The indicators of participation in the nonprobability cohort are uncorrelated with each other given the self-selection variables, i.e., $cov\left(\delta_i^{(c)}, \delta_j^{(c)} \mid x_i, x_j\right) = 0$ for $i \neq j$.

An independent reference probability-based survey sample $s_p$ of size $n_p$ is randomly selected from $FP$. The sample inclusion indicator, selection probability, and the corresponding sample weights are defined by $\delta_i^{(p)}$ (=1 if $i \in s_p$; 0 otherwise), $\pi_i^{(p)} = E_p\left(\delta_i^{(p)} \mid x_i\right)$, and $d_i = \frac{1}{\pi_i^{(p)}}$, respectively, where $E_p$ is with respect to the survey sample selection.

## 2.2. Existing logistic propensity weighting method

In this section, we first briefly introduce the existing RDW and CLW methods and discuss their pros and cons.

### 2.2.1 Rescaled design weight method (RDW)

Valliant and Dever [6, 7] assumed a logistic regression model for the participation rates $\pi_i^{(c)}(\gamma)$

$$\log\left\{\frac{\pi_i^{(c)}(\gamma)}{1 - \pi_i^{(c)}(\gamma)}\right\} = \gamma^T x_i, \text{ for } i \in FP, \tag{2.2.1}$$

where $\gamma$ is a vector of unknown parameters, and $x_i$ is a vector of covariates for $i \in FP$. To simplify the notation, we use $\pi_i^{(c)}$ below. They considered (implicitly) the population likelihood function of $\pi_i^{(c)}$ as

$$L(\gamma) = \prod_{i \in FP} \left\{\pi_i^{(c)}\right\}^{\delta_i^{(c)}} \left\{1 - \pi_i^{(c)}\right\}^{1 - \delta_i^{(c)}}, \tag{2.2.2}$$

Then, the log-likelihood function can be written as

$$\begin{aligned} l(\gamma) &= \sum_{i \in FP} \left[\delta_i^{(c)} \log \pi_i^{(c)} + \left\{1 - \delta_i^{(c)}\right\} \log\left\{1 - \pi_i^{(c)}\right\}\right] \\ &= \sum_{i \in s_c} \log \pi_i^{(c)} + \sum_{i \in FP - s_c} \log\left\{1 - \pi_i^{(c)}\right\}, \end{aligned} \tag{2.2.3}$$

where the set $FP - s_c$ represents the finite population units that are not self-selected into the nonprobability sample. Since $FP - s_c$ is not available in practice, the pseudo-loglikelihood function was constructed to estimate $l(\gamma)$ by

$$\tilde{l}^{RDW}(\pmb{\gamma}) = \sum_{i \in s_c} w_i^* \log \pi_i^{(c)} + \sum_{i \in s_p} w_i^* \log\{1 - \pi_i^{(c)}\} \tag{2.2.4}$$

where $w_i^* = \begin{cases} 1, & \text{for } i \in s_c \\ d_i \frac{\widehat{N}_p - n_c}{\widehat{N}_p}, & \text{for } i \in s_p \end{cases}$, with $\widehat{N}_p = \sum_{i \in s_p} d_i$ being the survey estimate of the target finite population size $N$. This leads to the total of the scaled weights across the probability sample units being $\sum_{i \in s_p} w_i^* = \widehat{N}_p - n_c$. The rationale for rescaling is to weight the survey sample to represent the complement of $s_c$ in the finite population, i.e., the set $FP - s_c$. Under the logistic regression model, the nonprobability sample participation rate $\pi_i^{(c)}$ for $i \in s_c$ can be estimated by fitting Model (2.2.1) to the combined sample of $s_c$ and *scaled-weighted* $s_p$ with scaled weights $w_i^*$, leading to the RDW estimates.

The RDW method has been shown to effectively reduce the bias of the naïve nonprobability sample estimates. However, the summand $\sum_{i \in FP - s_c} \log\{1 - \pi_i^{(c)}\}$ in (2.2.3) is not a fixed finite population total because units in the nonprobability sample $s_c$ are treated as being randomly observed. This leads to a bias as shown below.

Comparing the expectation of the population log-likelihood function $l(\pmb{\gamma})$ in (2.2.3) and the expectation of the pseudo log-likelihood $\tilde{l}^{RDW}(\pmb{\gamma})$ in (2.2.4), and letting $E(\cdot) = E_c E_p(\cdot)$ we have

$$E\{l(\pmb{\gamma})\} = \sum_{i \in FP} \pi_i^{(c)} \log \pi_i^{(c)} + \sum_{i \in FP} \{1 - \pi_i^{(c)}\} \log\{1 - \pi_i^{(c)}\}, \text{ and}$$

$$E\{\tilde{l}^{RDW}(\pmb{\gamma})\} = E_c E_p\{\tilde{l}^{RDW}(\pmb{\gamma})\}$$

$$= E_c\left[\sum_{i \in FP} \delta_i^{(c)} \log \pi_i^{(c)}\right] + E_p\left[\sum_{i \in FP} \delta_i^{(p)} \cdot \frac{\widehat{N}_p - n_c}{\widehat{N}_p} d_i \log\{1 - \pi_i^{(c)}\}\right]$$

$$\doteq \sum_{i \in FP} \pi_i^{(c)} \log \pi_i^{(c)} + \sum_{i \in FP} \left\{1 - \frac{n_c}{N}\right\} \log\{1 - \pi_i^{(c)}\}$$

by assuming $E_p(\widehat{N}_p) = N$. The difference of the two expectations, denoted by $\Delta_{RDW}$, can be written as

$$\Delta_{RDW} = E\{\tilde{l}^{RDW}(\gamma)\} - E\{l(\gamma)\} = \sum_{i \in FP} \left\{\frac{n_c}{N} - \pi_i^{(c)}\right\} \log\{1 - \pi_i^{(c)}\}$$

which, in general, is nonzero. Accordingly, the nonprobability sample participation rates estimated by solving for $\gamma$ in $\partial \tilde{l}^{RDW}(\gamma)/\partial \gamma = 0$ under Model (2.2.1) can be biased, unless either (i) the nonprobability sample units have small participation rates, i.e., both $n_c/N$ and $\pi_i^{(c)}$ are close to 0 for all $i \in FP$, in which case $\log\{1 - \pi_i^{(c)}\} \approx 0$, or (ii) all population units are equally likely to participate in the nonprobability sample, i.e. $\pi_i^{(c)} \equiv n_c/N$. In many practical applications, (i) will hold. For example, suppose that $n_c = 1000$ and the US population age 18 and over is the target population. The population size is approximately 210 million, so that $n_c/N \doteq 5 \times 10^{-6}$. If, instead, the population is for a small state like Wyoming where the 18+ population size is about 365,000, then $n_c/N \doteq 0.0027$. In both examples, with such small sampling fractions, all $\{\pi_i^{(c)}, i \in s_c\}$ should be near zero also.

### 2.2.2 CLW Method

Chen et al [4] proposed another IPSW method using the same likelihood function $L(\gamma)$ in (2.2.2), but rewriting the population log-likelihood as

$$l(\gamma) = \sum_{i \in s_c} \log \frac{\pi_i^{(c)}}{1 - \pi_i^{(c)}} + \sum_{i \in FP} \log\{1 - \pi_i^{(c)}\}. \tag{2.2.5}$$

In contrast to the RDW method, CLW estimated the population total of $\log\{1 - \pi_i^{(c)}\}$ by a weighted reference sample total and constructed the pseudo log-likelihood as

$$\tilde{l}^{CLW}(\gamma) = \sum_{i \in s_c} \log \frac{\pi_i^{(c)}}{1 - \pi_i^{(c)}} + \sum_{i \in s_s} d_i \log\{1 - \pi_i^{(c)}\}. \tag{2.2.6}$$

Under the same logistic regression model (2.2.1), the participation rate $\pi_i^{(c)}$ was estimated by solving the pseudo estimation equation

$$\tilde{S}(\gamma) = \frac{1}{N}\left(\sum_{i \in s_c} x_i - \sum_{i \in s_s} d_i \pi_i^{(c)} x_i\right) = 0, \tag{2.2.7}$$

derived from the pseudo log-likelihood (2.2.6). CLW proved that the resulting estimator of the finite population mean was design consistent when model (2.2.1) for the participation rates was correct.

In contrast to the RDW method, CLW does not require condition (i) or (ii) in the RDW method for unbiased estimation of participation rates $\{\pi_i^{(c)}, i \in s_c\}$. In the next section, we propose an adjusted logistic propensity (ALP) method, which corrects the bias in the RDW method. The proposed ALP method provides consistent estimators of finite population means and is as easy to implement as the RDW method.

## 2.3. Adjusted logistic propensity method (ALP)

The ALP method also aims to estimate the cohort sample participation rates $\{\pi_i^{(c)}, i \in s_c\}$ and use the inverse of estimated $\pi_i^{(c)}$ as the pseudo-weight for $i \in s_c$. As a computational device, we construct a pseudo-population of $s_c^* \cup FP$, where $s_c^*$ is a copy of $s_c$ that has the same joint distributions of covariates $x$ and outcome $y$ with the original $s_c$. The number of units in $s_c^* \cup FP$ is $n_c + N$. In the union of $s_c^* \cup FP$, $s_c^*$ and $s_c$ are treated as two different sets. We use $R_i$ to indicate the membership of $s_c^*$ in $s_c^* \cup FP$ (=1 if $i \in s_c^*$; 0 if $i \in FP$), and $p_i = P(R_i = 1) = P(i \in s_c^* \mid s_c^* \cup FP)$. Instead of directly modeling $\pi_i^{(c)}$ as in the RDW and CLW methods, we model $p_i$ as a function of $\pi_i^{(c)}$:

$$p_i = \frac{\pi_i^{(c)}}{1 + \pi_i^{(c)}}, \quad \text{or equivalently,} \quad \pi_i^{(c)} = \frac{p_i}{1 - p_i}. \tag{2.3.1}$$

The relationship between $p_i$ and $\pi_i^{(c)}$ follows because

$$\frac{p_i}{1 - p_i} = \frac{P(i \in s_c^* \mid s_c^* \cup FP)}{P(i \in FP \mid s_c^* \cup FP)} = \frac{P(i \in s_c)}{P(i \in FP)} = P(i \in s_c \mid FP) = \pi_i^{(c)} \tag{2.3.2}$$

since $s_c^*$ is a copy of $s_c$ and $P(i \in s_c^* \mid s_c^* \cup FP) = P(i \in s_c \mid s_c^* \cup FP)$. Notice that, derived from Formula (2.3.1),

$$p_i \leq \frac{1}{2},$$

since $\pi_i^{(c)} \leq 1$ and the equality holds only if $\pi_i^{(c)} = 1$, i.e., the $FP$ unit $i$ participates in the cohort with certainty. As illustrated by the examples at the end of section 2.2.1, requiring $p_i \leq 1/2$ is not unrealistic in typical applications because $\pi_i^{(c)}$ is generally quite small.

Suppose that $p_i$ can be modeled parametrically by $p_i = p(x_i; \boldsymbol{\beta}) = \text{expit}(\boldsymbol{\beta}^T x_i)$, where $\boldsymbol{\beta}$ is a vector of unknown model parameters. That is,

$$\log\left\{\frac{p_i}{1 - p_i}\right\} = \boldsymbol{\beta}^T x_i, \text{ for } i \in s_c^* \cup FP \tag{2.3.3}$$

Notice that $\boldsymbol{\beta}$, the coefficients in Model (2.3.3), differ from the coefficients $\boldsymbol{\gamma}$ in Model (2.2.1) because the two logistic regression models have different dependent variables. Based on (2.3.1), expression (2.3.3) implies that $\pi_i^{(c)}$ is being modeled as $\exp(\boldsymbol{\beta}^T x_i)$, which differs from the RDW/CLW model in (2.2.1) where $\pi_i^{(c)} = \exp(\boldsymbol{\gamma}^T x_i)/\{1 + \exp(\boldsymbol{\gamma}^T x_i)\}$. The corresponding "likelihood" function can be written as

$$L^*(\boldsymbol{\beta}) = \prod_{i \in s_c^* \cup FP} p_i^{R_i}(1 - p_i)^{(1 - R_i)}, \tag{2.3.4}$$

where $R_i$ indicates the membership of $s_c^*$ in $s_c^* \cup FP$ (=1 if $i \in s_c^*$; 0 if $i \in FP$). We put "likelihood" in quotes because $L^*(\boldsymbol{\beta})$ varies depending on which set of units is selected for $s_c^*$. This contrasts with the population likelihood in (2.2.2) which applies regardless of which sample is selected. Note that $L^*(\boldsymbol{\beta})$ in (2.3.4) is written as if the units are independent when they are not. This is a standard procedure in pseudo-MLE estimation, and the resulting parameter estimators remain design-consistent even when some units may be correlated due to, e.g., clustering (Binder[9]; Chambers & Skinner, sec. 6.3[10]). The quantity $L^*(\boldsymbol{\beta})$ should be viewed as motivation for developing the estimating equations given below in (2.3.7). The log-likelihood generated from $L^*(\boldsymbol{\beta})$ is

$$l^*(\boldsymbol{\beta}) = \sum_{i \in s_c^* \cup FP} \{R_i \cdot \log p_i + (1 - R_i) \log(1 - p_i)\}$$

$$= \sum_{i \in s_c^*} \log p_i + \sum_{i \in FP} \log(1 - p_i) \quad (2.3.5)$$

$$= \sum_{i \in FP} \delta_i^{(c)} \log p_i + \sum_{i \in FP} \log(1 - p_i)$$

Notice that the randomness of $L^*(\boldsymbol{\beta})$ and $l^*(\boldsymbol{\beta})$ comes from the cohort selection, i.e. $\delta_i^{(c)}$ in the first summand in the last line of (2.3.5). In reality, since the unit level information of $FP$ is unknown, we replace the second summand in (2.3.5) by a survey sample estimate, $\sum_{i \in s_p} d_i \log(1 - p_i)$. The pseudo log-likelihood is

$$\tilde{l}^*(\boldsymbol{\beta}) = \sum_{i \in FP} \delta_i^{(c)} \log p_i + \sum_{i \in FP} d_i \, \delta_i^{(p)} \log(1 - p_i), \quad (2.3.6)$$

where $\delta_i^{(c)}$ and $\delta_i^{(p)}$ are, respectively, the sample (self-)selection indicator for the cohort and survey sample defined in Section 2.1. The maximum pseudo-likelihood estimator $\widehat{\boldsymbol{\beta}}$ from (2.3.6) can be obtained by solving the pseudo-estimating equation $\tilde{S}^*(\boldsymbol{\beta}) = 0$, where

$$\tilde{S}^*(\boldsymbol{\beta}) = \frac{1}{N + n_c} \left\{ \sum_{i \in s_c} (1 - p_i) \boldsymbol{x}_i - \sum_{i \in s_p} d_i p_i \boldsymbol{x}_i \right\}. \quad (2.3.7)$$

Assuming that $p_i$ is bounded by 0 and $\frac{1}{2}$ implies that $\pi_i^{(c)}$ is automatically bounded by 0 and 1. Note that (2.3.7) has the same form as estimating equation (6) in CLW[4] if their function $h(\boldsymbol{x}_i, \boldsymbol{\theta})$ is set equal to $x_i\{1 + \pi_i^{(c)}\}^{-1}$; the same estimating equations were also studied earlier in Beaumont[11] and Kim and Kim[12].

The ALP estimator of $\mu$ is

$$\hat{\mu}^{ALP} = \frac{\sum_{i \in s_c} w_i^{ALP} y_i}{\sum_{i \in s_c} w_i^{ALP}}, \qquad (2.3.8)$$

where $w_i^{ALP} = 1/\pi_i^{(c)}(\hat{\boldsymbol{\beta}})$ for $i \in s_c$. Although $L^*(\boldsymbol{\beta})$ is not a standard likelihood, $\hat{\mu}^{ALP}$ is a consistent estimator of the population mean as shown in the theorem below.

We consider the following limiting process for the theoretical development.[4, 13] Suppose there is a sequence of finite populations $FP_k$ of size $N_k$, for $k = 1, 2, \cdots$. Cohort $s_{c,k}$ of size $n_{c,k}$ and survey sample $s_{p,k}$ of size $n_{p,k}$ are sampled from $FP_k$. The sequences of the finite population, the cohort and the survey sample have their sizes satisfy $\lim_{k \to \infty} \frac{n_{t,k}}{N_k} \to f_t$, where $t = c$ or $p$ and $0 < f_t \leq 1$ (regularity condition C1 in Appendix A). In the following the index $k$ is suppressed for simplicity.

**Theorem.** Consistency of ALP estimator of finite population mean (see Appendix B)

Under the regularity conditions A1-A3, and C1-C5 in Appendix A, and assuming logistic regression model (2.3.3) for $p_i$, the ALP estimate $\hat{\mu}^{ALP}$ is design consistent for $\mu$, in particular $\hat{\mu}^{ALP} - \mu = O_p(n_c^{-1/2})$, with the finite population variance

$$Var(\hat{\mu}^{ALP}) \doteq N^{-2} \sum_{i \in FP} p_i(1 - 2p_i) \left\{ \frac{(y_i - \mu)}{p_i} - \boldsymbol{b}^T \boldsymbol{x}_i \right\}^2 + \boldsymbol{b}^T \boldsymbol{D} \boldsymbol{b}, \qquad (2.3.9)$$

where $p_i = \text{expit}(\boldsymbol{\beta}^T \boldsymbol{x}_i)$, $\boldsymbol{b}^T = \{\sum_{i \in FP}(y_i - \mu)\boldsymbol{x}_i^T\}\{\sum_{i \in FP} p_i \boldsymbol{x}_i \boldsymbol{x}_i^T\}^{-1}$, and $\boldsymbol{D} = N^{-2} V_p\left(\sum_{i \in s_p} d_i p_i \boldsymbol{x}_i\right)$ is the design-based variance-covariance matrix under the probability sampling design for $s_p$.

In practice, the ALP estimator of a finite population mean can be obtained by three steps:

Step 1: Search for covariates $x$ available in both the cohort ($s_c$) and the reference survey sample ($s_p$) and combine the two samples. Assign $R_i = 1$ for $i \in s_c$ and $R_i = 0$ for $i \in s_p$ in the combined sample.

Step 2: Fit a logistic regression model for $p_i = P(R_i = 1)$ in the combined $s_c$ and weighted $s_p$, with the survey sample weights $\{d_i, i \in s_p\}$, and obtain the estimate $\hat{p}_i$ for $i \in s_c$.

Step 3: Estimate the finite population mean by Formula (2.3.8) with the ALP pseudo weight $w_i^{ALP} = \hat{p}_i/(1-\hat{p}_i)$ for $i \in s_c$.

Notice that Step 2 can be accomplished by any existing survey software, such as svyglm in survey package of R, svy:logit in Stata, and PROC SURVEYLOGISTIC in SAS. In addition to being easy to implement, the ALP estimator from (2.3.8) does not require conditions (i) or (ii), unlike RDW. Moreover, we prove that in large samples, $Var(\hat{\mu}^{ALP}) = O(n_c^{-1})$ is as or more efficient compared to $Var(\hat{\mu}^{CLW}) = O\left\{\min(n_p, n_c)^{-1}\right\}$, which depends on both the nonprobability and probability sample sizes (see Appendix C).

An alternative method would be to omit the odds transformation, which uses $p_i$ to approximate the participation rate $\pi_i^{(c)}$. Denote this method by FDW for full design weight, which contrasts to the scaling of the survey sample weights in the RDW method. Comparing the expectation of the population log-likelihood function $l(\gamma)$ in (2.2.3) and the expectation of the pseudo log-likelihood $\tilde{l}^*(\boldsymbol{\beta})$ in (2.3.5) with $\pi_i^{(c)}$ replacing $p_i$ by the FDW method, i.e., $\tilde{l}^*(\gamma) = \sum_{i \in s_c} \log \pi_i^{(c)} + \sum_{i \in s_p} d_i \log(1 - \pi_i^{(c)})$, we have their difference, denoted by $\Delta_{FDW}$, written as

$$\Delta_{FDW} = E\{\tilde{l}^*(\gamma)\} - E\{l(\gamma)\}$$

$$= \sum_{i \in FP} \pi_i^{(c)} \log \pi_i^{(c)} + \sum_{i \in FP} \log\{1 - \pi_i^{(c)}\} - \sum_{i \in FP} \pi_i^{(c)} \log \pi_i^{(c)}$$

$$-\sum_{i \in FP} \left(1 - \pi_i^{(c)}\right) \log\left\{1 - \pi_i^{(c)}\right\}$$

$$= \sum_{i \in FP} \pi_i^{(c)} \log\left\{1 - \pi_i^{(c)}\right\}.$$

The bias is zero only if $\pi_i^{(c)}$ for $i \in FP$ are all close to zero. Thus, the odds transformation step in ALP could be skipped if all nonprobability participation rates are extremely small; but, in general, that step is essential for unbiased estimation.

### 2.4. Variance estimation

Using the finite population variance formula (2.3.9), the first summand can be consistently estimated by

$$\{\widehat{N}^{(c)}\}^{-2} \sum_{i \in s_c} (1 - \hat{p}_i)(1 - 2\hat{p}_i) \left\{\frac{(y_i - \hat{\mu}^{ALP})}{\hat{p}_i} - \widehat{\boldsymbol{b}}^T \boldsymbol{x}_i\right\}^2, \qquad (2.4.1)$$

where $\hat{p}_i$ is the prediction for $i \in s_c$, $\widehat{N}^{(c)} = \sum_{i \in s_c} w_i^{ALP}$, and $\widehat{\boldsymbol{b}}^T = \{\sum_{i \in s_c}(y_i - \hat{\mu}^{ALP})\boldsymbol{x}_i^T\}\{\sum_{i \in s_c} \hat{p}_i \boldsymbol{x}_i \boldsymbol{x}_i^T\}^{-1}$. The second summand $\boldsymbol{b}^T \boldsymbol{D} \boldsymbol{b}$ is estimated by $\widehat{\boldsymbol{b}}^T \widehat{\boldsymbol{D}} \widehat{\boldsymbol{b}}$, where $\widehat{\boldsymbol{D}}$ is the survey design consistent variance estimator of $\boldsymbol{D}$. For example, under stratified multistage cluster sampling with $H$ strata and $a_h$ primary sampling units (PSUs) in stratum $h$ selected with replacement,

$$\widehat{\boldsymbol{D}} = \{\widehat{N}^{(p)}\}^{-2} \cdot \sum_{h=1}^{H} \frac{a_h}{a_h - 1} \sum_{l=1}^{a_h} (\boldsymbol{z}_l - \bar{\boldsymbol{z}})(\boldsymbol{z}_l - \bar{\boldsymbol{z}})^T, \qquad (2.4.2)$$

where $\widehat{N}^{(p)} = \sum_{i \in s_p} d_i$, $\boldsymbol{z}_l = \sum_{i \in s_p(hl)} d_i \hat{p}_i \boldsymbol{x}_i$ is the weighted PSU total for cluster $l$ in stratum $h$, $s_p(hl)$ is the set of sample elements stratum $h$ and cluster $l$, and $\bar{\boldsymbol{z}} = \frac{1}{a_h} \sum_l^{a_h} \boldsymbol{z}_l$ is the mean of the PSU totals in stratum $h$.

## 2.5. Scaling survey weights in the likelihood for the ALP Method

Unlike to the CLW method, the proposed ALP can flexibly scale the survey weights in estimating equation (2.3.7) to improve efficiency. For case-control studies, Scott & Wild [14] and Li et al [15] previously used the technique we propose below to reduce variances of estimates of relative risks when weights for cases and controls are substantially different. We multiply the second summand in $\tilde{S}^*(\boldsymbol{\beta})$ by a constant $\lambda$, say $\lambda = \left(\frac{n_c}{\sum_{i \in s_p} d_i}\right)$, so that the sum of the scaled survey weights ($\lambda d_i$) is $n_c$. Accordingly, the score function becomes

$$\tilde{S}_\lambda^*(\boldsymbol{\beta}) = \sum_{i \in s_c} (1 - p_i) \boldsymbol{x}_i - \lambda \sum_{i \in s_p} d_i p_i \boldsymbol{x}_i \qquad (2.5.1)$$

Solving $\tilde{S}_\lambda^*(\boldsymbol{\beta}) = 0$ for $\boldsymbol{\beta}$, and the resulting vector of estimates is denoted by $\widehat{\boldsymbol{\beta}}_\lambda = (\hat{\beta}_{0,\lambda}, \widehat{\boldsymbol{\beta}}_{1,\lambda})$, where $\hat{\beta}_{0,\lambda}$ is estimate of the intercept. Similar derivations to those in Scott & Wild [10] and Li et al [11] can be used to prove that $\widehat{\boldsymbol{\beta}}_{1,\lambda}$ is design-consistent with various efficiency gains, depending on the variability of survey weights versus the nonprobability sample weights (with implicit common value of 1). However, the estimate of the intercept $\hat{\beta}_{0,\lambda}$ can be badly biased with scaled weights. As a result, the estimate of participation rate $\exp(\widehat{\boldsymbol{\beta}}_\lambda^T \boldsymbol{x}_i)$ including $\hat{\beta}_{0,\lambda}$ would also be biased. The bias of $\hat{\beta}_{0,\lambda}$, however, would not affect the estimate of population mean because the scaled ALP-weighted mean, $\hat{\mu}^{ALP.S}$,

$$\hat{\mu}^{ALP.S} = \frac{\sum_{i \in s_c} w_i^{ALP.S} y_i}{\sum_{i \in s_c} w_i^{ALP.S}} = \frac{\sum_{i \in s_c} \exp^{-1}(\widehat{\boldsymbol{\beta}}_{1,\lambda}^T \boldsymbol{x}_i) y_i}{\sum_{i \in s_c} \exp^{-1}(\widehat{\boldsymbol{\beta}}_{1,\lambda}^T \boldsymbol{x}_i)}, \qquad (2.5.2)$$

depends on $\widehat{\boldsymbol{\beta}}_{1,\lambda}$, but not $\hat{\beta}_{0,\lambda}$.

It can be proved that $\hat{\mu}^{ALP.S}$ is a consistent estimator of the finite population mean, $\mu$. The Taylor linearization (TL) variance estimator of $\hat{\mu}^{ALP.S}$ can be obtained by substituting $w_i^{ALP}$, $\hat{\mu}^{ALP}$, $\widehat{\boldsymbol{\beta}}$, $\hat{p}_i$ and $d_i$ by $w_i^{ALP.S}$, $\hat{\mu}^{ALP.S}$, $\widehat{\boldsymbol{\beta}}_\lambda$, $\hat{p}_{i,\lambda} = \exp(\widehat{\boldsymbol{\beta}}_\lambda^T \boldsymbol{x}_i)$ and $\lambda d_i$, respectively, in Formulae (2.4.1) and (2.4.2). Details on the variance and the consistency of ALP.S are discussed in the dissertation by Wang.[16]

## 3. SIMULATIONS

### 3.1. Finite population generation and sample selection

We applied simulation setups similar to those in Chen et al.[4] In the finite population $FP$ of size $N = 500,000$, a vector of covariates $\mathbf{x}_i = (x_{1i}, x_{2i}, x_{3i}, x_{4i})^T$ was generated for $i \in FP$ where $x_{1i} = v_{1i}$, $x_{2i} = v_{2i} + 0.3x_{1i}$, $x_{3i} = v_{3i} + 0.2(x_{1i} + x_{2i})$, $x_{4i} = v_{4i} + 0.1(x_{1i} + x_{2i} + x_{3i})$, with $v_{1i} \sim Bernoulli(0.5)$, $v_{2i} \sim Uniform(0, 2)$, $v_{3i} \sim Exponential(1)$, and $v_{4i} \sim \chi^2(4)$. The variable of interest $y_i \sim Normal(\mu_i, 1)$, where $\mu_i = -x_{1i} - x_{2i} + x_{3i} + x_{4i}$ for $i \in FP$. The parameter of interest was the finite population mean $\mu = \frac{1}{N}\sum_{i \in FP} y_i = 3.97$.

The probability-based survey sample $s_p$ with the target sample size $n_p = 12,500$ (sampling fraction $f_p = 2.5\%$) was selected by Poisson sampling, with inclusion probability $\pi_i^{(p)} = (n_p \cdot q_i)/\sum_{i \in FP} q_i$ for $i \in FP$, where $q_i = const + x_{3i} + 0.03y_i$ with controlling for the variation of the survey weights, $1/\pi_i^{(p)}$. We set $const = -0.26$ so that $\max q_i / \min q_i = 20$.

As noted in section 2, the ALP and CLW methods do assume somewhat different models for the participation rate. Thus, it is interesting to check their performances both when their underlying models are correct and when the assumed participation rate models fail. The volunteer-based nonprobability sample $s_c$ (with a target sample size $n_c$) was also selected by Poisson sampling but with different inclusion probabilities $\pi_i^{(c)}$ for $i \in FP$. We considered two scenarios with different functional forms of $\pi_i^{(c)}$ so that the ALP (and FDW) or the CLW method had the true linear logistic regression propensity model in one scenario but not in the other. In Scenario 1, $\pi_i^{(c)} = \exp(\beta_0 + \boldsymbol{\beta}^T \mathbf{x}_i)$ was the specified participation rate for the $i^{th}$ population unit to be included into the nonprobability sample. The underlying true propensity model for ALP (and FDW) methods, shown in (2.3.3), was $\log\left\{\frac{p_i}{1-p_i}\right\} = \log\{\pi_i^{(c)}\} = \beta_0 + \boldsymbol{\beta}^T \boldsymbol{x}_i$, which implies $\log\left\{\frac{\pi_i^{(c)}}{1-\pi_i^{(c)}}\right\} = \beta_0 + \boldsymbol{\beta}^T \boldsymbol{x}_i - \log\{1 - \pi_i^{(c)}\}$. This model differs from the underlying linear model (2.2.1) assumed by the CLW method by the addition of the term $\log\{1 - \pi_i^{(c)}\}$. In Scenario 2, $\pi_i^{(c)} = \text{expit}(\gamma_0 + \boldsymbol{\gamma}^T \boldsymbol{x}_i)$ was specified so that $\log\left\{\frac{\pi_i^{(c)}}{1-\pi_i^{(c)}}\right\} = \gamma_0 + \boldsymbol{\gamma}^T \boldsymbol{x}_i$, which was the model (2.2.1) assumed by the CLW method. This model, however, implied that $\log\left\{\frac{p_i}{1-p_i}\right\} = \log\{\pi_i^{(c)}\} = \gamma_0 + \boldsymbol{\gamma}^T \boldsymbol{x}_i + \log\{1 - \pi_i^{(c)}\}$, which was different from the model assumed by the ALP and the FDW method (by the extra term $\log\{1 - \pi_i^{(c)}\}$). Hence, ALP and CLW estimates of the population mean are expected to be unbiased in one scenario but not the other since both methods assume a linear logistic propensity model. The biases of the FDW and RDW estimates, as measured by $\Delta_{FDW}$ and $\Delta_{RDW}$, depend on $\pi_i^{(c)}$, and go to 0 as $\pi_i^{(c)}$ approached 0. The biases become larger as $\pi_i^{(c)}$ increases in either scenario.

In both scenarios, the coefficients were set to be $\boldsymbol{\beta} = \boldsymbol{\gamma} = (0.18, 0.18, -0.27, -0.27)^T$. The parameters were chosen so that $0 < \pi_i^{(c)} < 1$ for all units $i \in FP$. The intercepts $\beta_0$ and $\gamma_0$ were also controlled so that the expected number of nonprobability sample units $E_c(n_c) = \sum_{FP} \pi_i^{(c)}$ was varied from 1,250, 2,500, 5,000, to 10,000 with the corresponding overall participation rate $f_c = \frac{E_c(n_c)}{N}$ being 0.5%, 5%, 10%, or 20%.

## 3.2. Evaluation Criteria

We examined the performance of five IPSW estimators of finite population mean $\mu$: (1)-(2) $\hat{\mu}^{ALP}$ and $\hat{\mu}^{ALP.S}$ described in Section 2.2-2.5; (3) $\hat{\mu}^{FDW}$ using weights from the ALP method omitting the odds transformation; (4) $\hat{\mu}^{CLW}$ proposed by Chen et al [4]; and (5) $\hat{\mu}^{RDW}$ proposed by Valliant &

Dever [6], compared with the naïve nonprobability sample mean ($\hat{\mu}^{Naive}$) that did not use weights, and the weighted nonprobability sample mean, $\hat{\mu}^{TW}$, with weights equal to the inverse of the true nonprobability sample inclusion probabilities. Note that $\hat{\mu}^{TW}$ is unavailable in practice because the true nonprobability sample inclusion probabilities are unknown. Relative bias (%RB), empirical variance ($V$), mean squared error (MSE) of the point estimates were used to evaluate the performance of the four IPSW point estimates, calculated by

$$\%\text{RB} = \frac{1}{B}\sum_{b=1}^{B} \frac{\hat{\mu}^{(b)} - \mu}{\mu} \times 100, \quad V = \frac{1}{B-1}\sum_{b=1}^{B}\left\{\hat{\mu}^{(b)} - \frac{1}{B}\sum_{b=1}^{B}\hat{\mu}^{(b)}\right\}^2, \quad \text{MSE} = \frac{1}{B}\sum_{b=1}^{B}\left\{\hat{\mu}^{(b)} - \mu\right\}^2,$$

where $B = 4,000$ is the number of simulation runs, $\hat{\mu}^{(b)}$ is one of the point estimates obtained from the $b$th simulated sample, and $\mu$ is the true finite population mean.

We also evaluated the variance estimates using the variance ratio (VR) and 95% confidence interval coverage probability (CP), which were calculated as

$$\text{VR} = \frac{\frac{1}{B}\sum_{b=1}^{B}\hat{v}^{(b)}}{V} \times 100, \text{ and } \text{CP} = \frac{1}{B}\sum_{b=1}^{B} I\left(\mu \in CI^{(b)}\right),$$

where $\hat{v}^{(b)}$ is the proposed analytical variance estimate in simulated sample $b$, and $CI^{(b)} = \left(\hat{\mu}^{(b)} - 1.96\sqrt{\hat{v}^{(b)}},\ \hat{\mu}^{(b)} + 1.96\sqrt{\hat{v}^{(b)}}\right)$ is the 95% confidence interval from the $b$-th simulated sample.

### 3.3. Results

Table 1 presents simulation results for the seven nonprobability sample estimators of the finite population mean. The naïve estimator $\hat{\mu}^{Naive}$ that ignored the underlying sampling scheme had relative biases ranging from -36.5% to -42.8% while the true weighted nonprobability sample estimator, $\hat{\mu}^{TW}$, was approximately unbiased in all scenarios. The variance of $\hat{\mu}^{Naive}$ was much smaller than that of the other estimators, but its bias caused the MSE to be extremely high (not reported).

Consistent with the bias theory in section 2, the RDW point estimator $\hat{\mu}^{RDW}$ and the FDW point estimator $\hat{\mu}^{FDW}$ were approximately unbiased when $\pi_i^{(c)}$ was small for all $i \in FP$ and the overall participation rate $f_c = \frac{1}{N}\sum_{i \in FP} \pi_i^{(c)}$ was low, but more biased as $f_c$ increased. The coverage probabilities decreased correspondingly.

As expected, the ALP estimators $\hat{\mu}^{ALP}$ and $\hat{\mu}^{ALP.S}$ (or the CLW estimator $\hat{\mu}^{CLW}$) consistently provided unbiased point estimators in the scenarios where they were expected to be unbiased, i.e., scenario 1 for $\hat{\mu}^{ALP}$ and $\hat{\mu}^{ALP.S}$, and scenario 2 for $\hat{\mu}^{CLW}$. When the underlying model was incorrect for an estimator, biases occurred. For example, the relative biases of $\hat{\mu}^{CLW}$ in scenario 1 were 0.05%, 1.29%, 2.94%, and 7.80% as $f_c$ increased from 0.5%, 5%, 10%, to 20%, respectively. In scenario 2, the corresponding relative biases for $\hat{\mu}^{ALP}$ are -0.19%, -1.03%, -1.81%, and -2.85%.

Consistent with the theory in Section 2, the ALP estimator $\hat{\mu}^{ALP}$ was more efficient than $\hat{\mu}^{CLW}$ with consistently smaller empirical variances in all scenarios, especially when the nonprobability cohort size was much larger than the probability sample size. Among all considered methods, $\hat{\mu}^{ALP.S}$ was approximately unbiased with the smallest variance under Scenario 1 of the correct model. Under Scenario 2 of a misspecified model, $\hat{\mu}^{ALP.S}$ was biased but most efficient, and therefore achieved smallest MSE.

The variance estimators for $\hat{\mu}^{ALP}$, $\hat{\mu}^{ALP.S}$ and $\hat{\mu}^{CLW}$ performed very well (with VR's near 1), providing coverage probabilities close to the nominal level under the correct propensity models when $f_c$ was large. The lower coverage of the nominal level (about 88%) when $f_c = 0.5\%$ was due to the small sample bias with skewed distributions of underlying sampling weights in the selected nonprobability sample.

## 4. REAL DATA EXAMPLE

We use the same data example as Wang et al [6] for illustration purposes. We estimated prospective 15-year all-cause, all-cancer, and heart disease mortality rates for adults in the US using the adult household interview part of The Third U.S. National Health and Nutrition Examination Survey (NHANES III) III conducted in 1988-1994, with sample size $n_c = 20{,}050$. We ignored all complex design features of NHANES III and treated it as a nonprobability sample. The coefficient of variation (CV) of sample weights is 125%, indicating highly variable selection probabilities, and thus low representativeness of the unweighted sample. For estimating mortality rates, we approximated that the entire sample of NHANES III was randomly selected in 1991 (the midpoint of the data collection time period).

For the reference survey, we used 1994 U.S. National Health Interview Survey (NHIS) respondents to the supplement for monitoring achievement of the Healthy People Year 2000 objectives. Adults aged 18 and older are included (sample size $n_p = 19{,}738$). The 1994 NHIS used a multistage stratified cluster sample design with 125 strata and 248 pseudo-PSUs. [17,18] We collapsed strata with only one PSU with the next nearest stratum for variance estimation purposes. [19] Both samples of NHANES III and NHIS were linked to National Death Index (NDI) for mortality, allowing us to quantify the relative bias of unweighted NHANES estimates, assuming the NHIS estimates as the gold standard. Notice that the mortality information was obtained by statistical linkage between the survey sample and NDI, [20] but not responses from the questionnaires. The all-cancer and heart-disease mortality were classified according to National Center for Health Statistics death code. [21, 22]

The usage of NHANES III as the "nonprobability cohort" has several advantages for illuminating the performance of the propensity weighting methods. The "nonprobability sample" and the reference survey sample have approximately the same target population, data collection mode, and similar questionnaires. This ensures that the pseudo-weighted "nonprobability sample" could potentially represent the target population, and thus enables us to characterize the performance of the propensity weighting methods in real data.

The distributions of selected common covariates and variables of interests in the two samples are presented in Table 2. As expected, the variables in the weighted samples of NHANES and 1994 NHIS have very close distributions because both weighted samples represent approximately the same finite population. In contrast, covariates distribute quite differently in the *unweighted* NHANES from the weighted samples, especially for design variables such as age, race/ethnicity, poverty, and region, which leads to large biases in mortality rates estimated from the *unweighted* NHANES.

The propensity model included main effects of common demographic characteristics (age, sex race/ethnicity, region, and marital status), socioeconomic status (education level, poverty, and household income), tobacco usage (smoking status, and chewing tobacco), health variables (body mass index [BMI], and self-reported health status), and a quadratic term for age. Appendix D shows the final propensity models for the five considered methods.

To evaluate the performance of the five PS-based methods, we used relative difference from the NHIS estimate $\%\text{RD} = \frac{\hat{\mu} - \hat{\mu}^{NHIS}}{\hat{\mu}^{NHIS}} \times 100$, TL variance estimate ($V$), and estimated MSE $= (\hat{\mu} - \hat{\mu}^{NHIS})^2 + V$, which treated the NHIS estimate as truth. Table 3 shows that the naïve NHANES III estimate of overall mortality was ~52% biased from the NHIS estimate because older people who have higher mortalities were oversampled (Table 2). All five IPSW methods substantially reduced the bias from the naïve estimate. Consistent with the simulation results, the ALP, FDW, RDW, and CLW method yielded close estimates when the sample fraction of the nonprobability sample was small ($\hat{f}_c = \frac{n_c}{\hat{N}_p} = 1.06 \times 10^{-4}$ calculated from Table 2). The ALP.S method, by scaling the NHIS sample weights in propensity estimation, reduced more bias than the other methods, and was more efficient. Therefore, the ALP.S estimate had the smallest MSE. The results for inference of all-cancer mortality had the similar pattern as the results for all-cause mortality. All pseudo-weighting methods removed most bias of the naïve NHANES estimate (with %RD= -3.21%~2.07% reduced from 24.68%). In contrast, for heart-disease mortality, all pseudo-weighting methods were substantially less biased than the naïve estimate, with %RD= 42.58%~57.78% reduced from 133.66%, but the alternative estimators still had undesirably large biases themselves. The bias reduction is not as much as that for all-cancer or all-cause mortality, and this may be due to the omission of important predictors of having heart disease and of being observed in the nonprobability sample in the propensity model.

## 5. DISCUSSION

This paper proposed adjusted logistic propensity weighting methods for population inference using nonprobability samples. The proposed ALP method corrects the bias in the rescaled design weight method (RDW[7]) by formulating the problem in an innovative way. As does the RDW method, the proposed ALP method retains the advantage of easy implementation by fitting a propensity model with survey weights in ready-to-use software. The proposed ALP estimators are design consistent if the assumed model for participation rate is correct. Taylor linearization variance estimators for ALP estimates are derived. Consistency of the ALP finite population mean estimators was proved theoretically and evaluated numerically.

A primary competitor to ALP is the CLW estimator developed by Chen, et al.[4] If the nonprobability cohort is a small fraction of the population, ALP and CLW are very similar, although ALP does have computational advantages regardless of the size of the sampling fraction. As the sampling fraction increases, ALP and CLW become more distinct. Both ALP and CLW methods fit a propensity model to the combined nonprobability sample and a weighted survey sample. Highly variable weights in the combined sample can lead to low efficiency of the estimated propensity model coefficients. Therefore, the variances of the ALP and the CLW estimators of the finite population means can be large in some applications. However, the proposed ALP is proved analytically and numerically to have a variance that is less than or equal to that of the CLW method regardless of whether the propensity model underlying ALP is correct.

An alternative ALP with scaled weights produces consistent propensity estimates and further improves efficiency as shown in the simulation and the real data example. The CLW estimator with the scaled survey weights, albeit more efficient, is biased (simulation results not shown).

An important point is that ALP and CLW methods assume somewhat different logistic regression models for propensity score estimation. Propensity is defined as $\pi_i^{(c)} = \exp(\boldsymbol{\beta}^T \boldsymbol{x}_i)$ by ALP in expression (2.3.2) and $\pi_i^{(c)} = \exp(\boldsymbol{\gamma}^T \boldsymbol{x}_i)/\{1 + \exp(\boldsymbol{\gamma}^T \boldsymbol{x}_i)\}$ by CLW in expression (2.2.2). Model diagnostics should be developed to select which propensity model is more appropriate for a given data set and will be the focus of our future research.

Both ALP and CLW are inverse-propensity-score-weighting methods that directly use (functions of) the propensity score to estimate the cohort participation rate. They can be sensitive to propensity model misspecification (e.g., missing interaction terms in the fitted propensity model) due to inaccurate estimates of participation rates. In contrast, propensity-score-based matching methods (not included in this study) may be more robust to the model misspecification, because they use propensity scores to measure the similarity between survey and cohort sample units, and distribute survey sample weights to the cohort based on their similarity. Examples of matching methods are propensity-score adjustment by subclassification[23], propensity-score-based kernel weighting methods[5, 6, 16] and River's matching method[24].

There are a number of shortcomings associated with the estimation of propensity scores using logistic regression. First, the logistic model is susceptible to model misspecification, requiring assumptions regarding correct variable selection and functional form, including the choice of polynomial terms and multiple-way interactions. If any of these assumptions are incorrect, propensity score estimates can be biased, and balance may not be achieved when conditioning on the estimated PS. Second, implementing a search routine for model specification, such as repeatedly fitting logistic regression models while in/excluding predictor variables, interactions or transformations of variables can be computationally infeasible or suboptimal. In this context, parametric regression can be limiting in terms of possible model structures that can be searched over, particularly when many potential predictors are present (high dimensional data). Various machine learning methods for estimating the propensity score that incorporate survey weights will also be our future research interest.

**DATA AVALIABILITY STATEMENT**

The data that support the findings of this study are available on request from the corresponding author.

**REFERENCE**

1. Collins R. What makes UK Biobank special. Lancet. 2012; 379(9822):1173-4.
2. Fry A, Littlejohns TJ, Sudlow C, Doherty N, Adamska L, Sprosen T, Collins R, Allen NE. Comparison of sociodemographic and health-related characteristics of UK Biobank participants with those of the general population. American journal of epidemiology. 2017; 186(9):1026-34.

3. Elliott MR, Valliant R. Inference for nonprobability samples. Statistical Science. 2017; 32(2):249-64.

4. Chen Y, Li P, Wu C. Doubly Robust Inference With Nonprobability Survey Samples. Journal of the American Statistical Association. 2019; 2011-2021.

5. Wang L., Graubard B.I., Katki H, Li Y. Improving external validity of epidemiologic cohort Analyses: a kernel weighting approach. Journal of the Royal Statistical Society: Series A (Statistics in Society). 2020; DO-10.1111/rssa.12564.

6. Wang, L., Graubard, B. I., Katki, H. A., & Li, Y. (2020). Efficient and Robust Propensity-Score-Based Methods for Population Inference using Epidemiologic Cohorts. *arXiv preprint arXiv:2011.14850*.

7. Valliant R, Dever JA. Estimating propensity adjustments for volunteer web surveys. Sociological Methods & Research. 2011;40(1):105-37.

8. Valliant R. Comparing alternatives for estimation from nonprobability samples. Journal of Survey Statistics and Methodology. 2020;8(2):231-63.

9. Binder, D.A. On the Variances of Asymptotically Normal Estimators from Complex Surveys International Statistical Review. 51(3): 279-292.

10. Chambers, R.A., Skinner, C.J. Analysis of Survey Data. New York: Wiley. 2003.

11. Beaumont, J.-F. Calibrated Imputation in Surveys Under a Quasi-Model-Assisted Approach. Journal of the Royal Statistical Society, Series B. 2005; 67(3): 445–458.

12. Kim, J. K., and Kim, J. J. Nonresponse Weighting Adjustment Using Estimated Probability. The Canadian Journal of Statistics. 2007; 35(4): 501–514.

13. Krewski, D., Rao, J.N. Inference from stratified samples: properties of the linearization, jackknife and balanced repeated replication methods. The Annals of Statistics, 1981;9(5):1010-9.

14. Scott AJ, Wild CJ. Fitting logistic models under case-control or choice based sampling. Journal of the Royal Statistical Society: Series B (Methodological). 1986;48(2):170-82.

15. Li Y, Graubard BI, DiGaetano R. Weighting methods for population-based case–control studies with complex sampling. Journal of the Royal Statistical Society: Series C (Applied Statistics). 2011; 60(2):165-85.

16. Wang L. Improving external validity of epidemiologic analyses by incorporating data from population-based surveys. Doctoral dissertation, University of Maryland, College Park; 2020. https://drum.lib.umd.edu/handle/1903/26125.

17. Massey JT. Design and estimation for the national health interview survey, 1985-94. US Department of Health and Human Services, Public Health Service, Centers for Disease Control, National Center for Health Statistics; 1989.

18. Ezzati TM, Massey JT, Waksberg J, Chu A, Maurer KR. Sample design: Third National Health and Nutrition Examination Survey. Vital and health statistics. Series 2, Data evaluation and methods research. 1992; (113):1-35.

19. Hartley HO, Rao JN, Kiefer G. Variance estimation with one unit per stratum. Journal of the American Statistical Association. 1969; 64(327):841-51.

20. National Center for Health Statistics. National Death Index user's guide. Hyattsville, MD, 2013. Available at the following address:

    https://www.cdc.gov/nchs/data/ndi/ndi_users_guide.pdf

21. National Center for Health Statistics Data Linkage Public-Use Linked Mortality File Data Dictionary. Hyattsville, MD, 2015. Available at the following address https://www.cdc.gov/nchs/data/datalinkage/public-use-2015-linked-mortality-files-data-dictionary.pdf

22. National Center for Health Statistics Data Linkage Underlying and Multiple Cause of Death Codes. Hyattsville, MD, 2018. Available at the following address https://www.cdc.gov/nchs/data/datalinkage/underlying_and_multiple_cause_of_death_codes.pdf

23. Lee S, Valliant R. Estimation for volunteer panel web surveys using propensity score adjustment and calibration adjustment. Sociological Methods & Research. 2009; 37(3):319-43.

24. Rivers D. Sampling for web surveys. Paper presented at the Joint Statistical Meetings, Section on Survey Research Methods. Salt Lake City, Utah, 2007.

**SUPPORTING INFORMATION**

Additional supporting information may be found online in the Supporting Information section at the end of the article.

Table 1 Results from 4,000 simulated survey samples and nonprobability samples with low to high participation rates under various propensity score models

| | Scenario 1 True propensity model for ALP | | | | | Scenario 2 True propensity model for CLW | | | | |
|---|---|---|---|---|---|---|---|---|---|---|
| | %RB | $V$ ($\times 10^5$) | VR | MSE ($\times 10^5$) | $CP^5$ | %RB | $V$ ($\times 10^5$) | VR | MSE ($\times 10^5$) | CP |
| $f_c = 0.5\%$ | | | | | | | | | | |
| $\hat{\mu}^{Naive}$ | -42.76 | 0.22 | 0.99 | | | -42.61 | 0.22 | 1.00 | | |
| $\hat{\mu}^{TW}$ | -0.13 | 4.38 | 0.93 | 4.39 | 0.90 | -0.12 | 4.38 | 0.93 | 4.38 | 0.90 |
| $\hat{\mu}^{RDW}$ | -0.29 | 3.73 | 0.93 | 3.75 | 0.87 | -0.40 | 3.63 | 0.93 | 3.66 | 0.87 |
| $\hat{\mu}^{FDW}$ | -0.28 | 3.73 | 0.93 | 3.75 | 0.87 | -0.40 | 3.63 | 0.93 | 3.66 | 0.87 |
| $\hat{\mu}^{ALP}$ | -0.07 | 3.70 | 0.93 | 3.77 | 0.88 | -0.19 | 3.66 | 0.93 | 3.67 | 0.88 |
| $\hat{\mu}^{CLW}$ | 0.05 | 3.87 | 0.93 | 3.87 | 0.89 | -0.07 | 3.76 | 0.93 | 3.76 | 0.88 |
| $\hat{\mu}^{ALP.S}$ | -0.11 | 3.54 | 0.92 | 3.54 | 0.87 | -0.21 | 3.45 | 0.92 | 3.45 | 0.87 |
| $f_c = 5\%$ | | | | | | | | | | |
| $\hat{\mu}^{Naive}$ | -42.74 | 0.02 | 0.99 | | | -41.21 | 0.02 | 1.01 | | |
| $\hat{\mu}^{TW}$ | -0.04 | 0.50 | 0.98 | 0.50 | 0.92 | -0.02 | 0.46 | 1.00 | 0.47 | 0.93 |
| $\hat{\mu}^{RDW}$ | -2.15 | 0.56 | 1.00 | 1.29 | 0.66 | -3.05 | 0.43 | 1.01 | 1.89 | 0.45 |
| $\hat{\mu}^{FDW}$ | -2.05 | 0.57 | 1.00 | 1.23 | 0.68 | -2.95 | 0.43 | 1.01 | 1.81 | 0.47 |
| $\hat{\mu}^{ALP}$ | -0.01 | 0.62 | 1.00 | 0.62 | 0.94 | -1.03 | 0.47 | 1.01 | 0.64 | 0.85 |
| $\hat{\mu}^{CLW}$ | 1.29 | 0.84 | 1.00 | 1.10 | 0.95 | 0.01 | 0.61 | 1.01 | 0.61 | 0.94 |
| $\hat{\mu}^{ALP.S}$ | -0.05 | 0.45 | 1.00 | 0.45 | 0.92 | -0.63 | 0.35 | 1.02 | 0.41 | 0.86 |
| $f_c = 10\%$ | | | | | | | | | | |
| $\hat{\mu}^{Naive}$ | -42.74 | 0.01 | 1.11 | | | -39.65 | 0.01 | 1.11 | | |
| $\hat{\mu}^{TW}$ | -0.01 | 0.25 | 1.02 | 0.25 | 0.94 | -0.01 | 0.22 | 1.01 | 0.22 | 0.94 |
| $\hat{\mu}^{RDW}$ | -4.25 | 0.34 | 1.00 | 3.20 | 0.17 | -5.62 | 0.22 | 0.99 | 5.20 | 0.02 |
| $\hat{\mu}^{FDW}$ | -3.87 | 0.35 | 1.00 | 2.71 | 0.24 | -5.28 | 0.22 | 0.99 | 4.62 | 0.03 |
| $\hat{\mu}^{ALP}$ | 0.01 | 0.42 | 1.00 | 0.42 | 0.95 | -1.81 | 0.27 | 0.99 | 0.79 | 0.65 |
| $\hat{\mu}^{CLW}$ | 2.94 | 0.80 | 1.00 | 2.16 | 0.76 | 0.03 | 0.42 | 0.99 | 0.42 | 0.95 |
| $\hat{\mu}^{ALP.S}$ | -0.03 | 0.27 | 1.02 | 0.27 | 0.94 | -0.86 | 0.18 | 1.01 | 0.29 | 0.81 |
| $f_c = 20\%$ | | | | | | | | | | |
| $\hat{\mu}^{Naive}$ | -42.75 | 0.00 | 1.26 | | | -36.50 | 0.01 | 1.21 | | |
| $\hat{\mu}^{TW}$ | -0.02 | 0.15 | 0.93 | 0.15 | 0.93 | -0.02 | 0.11 | 0.96 | 0.11 | 0.93 |
| $\hat{\mu}^{RDW}$ | -8.58 | 0.21 | 0.95 | 11.83 | 0.00 | -9.59 | 0.10 | 0.97 | 14.60 | 0.00 |
| $\hat{\mu}^{FDW}$ | -7.15 | 0.23 | 0.96 | 8.29 | 0.01 | -8.51 | 0.11 | 0.98 | 11.53 | 0.00 |
| $\hat{\mu}^{ALP}$ | 0.00 | 0.32 | 0.96 | 0.32 | 0.95 | -2.85 | 0.16 | 0.98 | 1.44 | 0.19 |
| $\hat{\mu}^{CLW}$ | 7.80 | 1.67 | 0.92 | 11.27 | 0.06 | 0.01 | 0.33 | 0.98 | 0.33 | 0.95 |
| $\hat{\mu}^{ALP.S}$ | -0.03 | 0.19 | 0.96 | 0.19 | 0.94 | -1.02 | 0.10 | 1.00 | 0.27 | 0.73 |

Table 2 Distribution of selected common variables in NIH-AARP and NHIS

| Variable | | NHIS 1994 | | NHANES III | |
|---|---|---|---|---|---|
| Total Count | | $n_s = 19738$  $\widehat{N}_s = 189608549$ | | $n_c = 20050$  $\widehat{N}_s = 187647206$ | |
| | | % | Weighted % | % | Weighted % |
| Age Group | 18-24 years | 10.5 | 13.3 | 15.8 | 15.8 |
| | 25-44 years | 42.9 | 43.7 | 35.4 | 43.7 |
| | 45-64 years | 26.1 | 26.6 | 22.6 | 24.6 |
| | 65 years and older | 20.5 | 16.4 | 26.2 | 16.0 |
| Race | NH-White | 76.1 | 75.9 | 42.3 | 76.0 |
| | NH-Black | 12.6 | 11.2 | 27.4 | 11.2 |
| | Hispanic | 8.0 | 9.0 | 28.9 | 9.3 |
| | NH-Other | 3.3 | 4.0 | 1.5 | 3.5 |
| Region | Northeast | 20.7 | 20.5 | 14.6 | 20.8 |
| | Midwest | 26.1 | 25.1 | 19.2 | 24.1 |
| | South | 31.5 | 32.5 | 42.7 | 34.3 |
| | West | 21.6 | 21.9 | 23.5 | 20.9 |
| Poverty | No | 79.1 | 82.3 | 67.9 | 80.3 |
| | Yes | 13.1 | 10.6 | 21.4 | 12.1 |
| | Unknown | 7.8 | 7.0 | 10.7 | 7.6 |
| Education | Lower than High school | 20.1 | 19.1 | 42.5 | 26.6 |
| | High School/Some College | 58.7 | 59.6 | 45.9 | 54.1 |
| | College or higher | 21.2 | 21.3 | 11.6 | 19.3 |
| Health Status | Excellent/Very good | 60.5 | 62.0 | 39.0 | 51.6 |
| | Good | 25.7 | 25.7 | 35.9 | 32.7 |
| | Fair/Poor | 13.8 | 12.3 | 25.1 | 15.7 |
| Mortality | All-Cause | 20.8 | 17.6 | 26.7 | 17.1 |
| | Heart-Disease | 9.43 | 5.69 | 4.95 | 4.04 |
| | All-Cancer | 5.57 | 4.11 | 5.10 | 4.47 |

Table 3. Relative difference (%RD) of all-cause 15-year mortality estimates from the NHIS estimate with estimated variance ($V$) and mean squared error (MSE)

| Mortality | Method | Estimate (%) | %RD | $V \, (\times 10^5)$ | MSE $(\times 10^5)$ |
|---|---|---|---|---|---|
| **All cause** | NHIS | 17.6 | | | |
| | Naïve | 26.7 | 52.16 | | |
| | ALP | 18.6 | 6.08 | 1.87 | 13.27 |
| | FDW | 18.6 | 6.08 | 1.87 | 13.28 |
| | RDW | 18.6 | 6.08 | 1.87 | 13.28 |
| | CLW | 18.6 | 6.07 | 1.87 | 13.24 |
| | ALP.S | 17.2 | -2.05 | 1.08 | 2.37 |
| **All cancer** | NHIS | 4.5 | | | |
| | Naïve | 5.6 | 24.68 | | |
| | ALP | 4.6 | 2.07 | 0.38 | 0.46 |
| | FDW | 4.6 | 2.07 | 0.38 | 0.46 |
| | RDW | 4.6 | 2.07 | 0.38 | 0.46 |
| | CLW | 4.6 | 2.06 | 0.37 | 0.46 |
| | ALP.S | 4.3 | -3.21 | 0.32 | 0.53 |
| **Heart disease** | NHIS | 4.0 | | | |
| | Naïve | 9.4 | 133.66 | | |
| | ALP | 6.4 | 57.78 | 0.50 | 54.88 |
| | FDW | 6.4 | 57.78 | 0.50 | 54.90 |
| | RDW | 6.4 | 57.78 | 0.50 | 54.90 |
| | CLW | 6.4 | 57.77 | 0.50 | 54.87 |
| | ALP.S | 5.8 | 42.58 | 0.33 | 29.86 |

# Supporting Document for Adjusted Logistic Propensity Weighting Methods for Population Inference using Nonprobability Volunteer-Based Epidemiologic Cohorts

By Lingxiao Wang, Richard Valliant, and Yan Li

## A. Regularity Conditions

**C1** The finite population size $N$, the cohort sample sizes $n_c$, and survey sample size $n_s$ satisfy $\lim_{N\to\infty, n_c\to\infty} n_c/N = f_c \in (0,1)$, and $\lim_{N\to\infty, n_p\to\infty} n_p/N = f_p \in (0,1)$.

**C2** There exist constants $c_1$ and $c_2$ such that $0 < c_1 \leq N\pi_i^{(c)}/n_c \leq c_2$, and $0 < c_1 \leq N\pi_i^{(p)}/n_p \leq c_2$ for all units $i \in F$.

**C3** The finite population ($FP$) and the sample selection for $s_s$ satisfy $N^{-1}\sum_{i \in s_p} d_i r_i - N^{-1}\sum_{i \in FP} r_i = O_p(n_p^{-1/2})$, where $r_i$ includes $x_i$ and $y_i$ where the order in probability is with respect to the probability sampling mechanism used to select $s_p$ and $d_i = 1/\pi_i^{(p)}$.

**C4** The $FP$ and the $p_i$'s satisfy $N^{-1}\sum_{i \in FP} y_i^2 = O(1)$, $N^{-1}\sum_{i \in FP} \|x_i\|^3 = O(1)$, $N^{-1}\sum_{i \in FP} p_i x_i x_i^T = O(1)$ being a positive definite matrix.

**C5** The cohort participation and the survey sample selection satisfy $Cov\left(\delta_i^{(c)}, \delta_j^{(p)}\right) = 0$ for $i, j \in FP$.

Conditions **C1** – **C3** are regularly used in practice. Under **C1**, sample fractions of the nonprobability and probability sample are bounded. Condition **C2** indicates the (implicit) sample weights of nonprobability and probability sample units are bounded, i.e., $\pi_i^{(c)} = O\left(\frac{n_c}{N}\right)$ and $\pi_i^{(p)} = O\left(\frac{n_p}{N}\right)$, and the inclusion probabilities for the nonprobability and probability samples do not differ in terms of order of magnitude from simple random sampling. Condition **C3** guarantees consistency of the Horvitz-Thompson estimators obtained from the probability sample. Condition **C4** is the typical finite moment conditions to validate Taylor series expansions. Condition **C5** requires that selection of the nonprobability and the probability samples be independent, which simplifies the asymptotic variance calculation.

## B. Proof of Theorem

We consider the following limiting process (Krewski & Rao, 1981; Chen, Li &Wu, 2019).

Suppose there is a sequence of finite populations $FP_k$ of size $N_k$, for $k = 1, 2, \cdots$. Cohort $s_{c,k}$ of size $n_{c,k}$ and survey sample $s_{p,k}$ of size $n_{p,k}$ are sampled from each $FP_k$. The sequences of the finite population, the cohort and the survey sample have their sizes satisfy $\lim_{k \to \infty} \frac{n_{t,k}}{N_k} \to f_t$ where $t = c$ or $p$ and $0 < f_t \leq 1$ (regularity condition C1 in Appendix A). In the following the index $k$ is suppressed for simplicity.

Let $\boldsymbol{\eta}^T = (\mu, \boldsymbol{\beta}^T)$. The ALP estimate of the finite population mean, $\hat{\mu}^{ALP}$, given in expression (2.3.8) in the main text, along with the estimates of propensity model parameters, $\widehat{\boldsymbol{\beta}}$ (solution of $\tilde{S}^*(\boldsymbol{\beta}) = 0$ in estimating equation (2.3.7) in the main text), can be combined as $\widehat{\boldsymbol{\eta}}^T = (\hat{\mu}^{ALP}, \widehat{\boldsymbol{\beta}}^T)$, which is the solution to the joint pseudo estimating equations

$$\Phi(\boldsymbol{\eta}) = \begin{pmatrix} U(\mu) = \frac{1}{N} \sum_{i \in FP} \delta_i^{(c)} \widetilde{w}_i (y_i - \mu) \\ \tilde{S}^*(\boldsymbol{\beta}) = \frac{1}{N + n_c} \sum_{i \in FP} \delta_i^{(c)} (1 - p_i) x_i - \frac{1}{N + n_c} \sum_{i \in FP} \delta_i^{(p)} d_i p_i x_i \end{pmatrix} = \mathbf{0}, \quad (B.1)$$

where $\widetilde{w}_i = 1/\pi_i^{(c)} = (1 - p_i)/p_i$. Under the joint randomization of the propensity model (i.e., self-selection of $s_c$) and the sampling design of $s_s$, we have $E\{\Phi(\boldsymbol{\eta}_0)\} = \mathbf{0}$, where $\boldsymbol{\eta}_0^T = (\mu_0, \boldsymbol{\beta}_0^T)$ with $\mu_0$ and $\boldsymbol{\beta}_0$ being the true value of $\mu$ and $\boldsymbol{\beta}$ respectively. The consistency of $\widehat{\boldsymbol{\eta}}$ follows similar arguments to those in Chen, Li & Wu (2019) (which cited Section 3.2 of Tsiatis (2007)). Under the conditions **C1**-**C4**, we have $\Phi(\widehat{\boldsymbol{\eta}}) = \mathbf{0}$ By applying the first-order Taylor expansion, we have

$$\widehat{\boldsymbol{\eta}} - \boldsymbol{\eta}_0 \doteq [E\{\phi(\boldsymbol{\eta}_0)\}]^{-1} \Phi(\boldsymbol{\eta}_0), \quad (B.2)$$

where $E\{\phi(\boldsymbol{\eta})\} = E\left\{\frac{\partial \Phi(\boldsymbol{\eta})}{\partial \boldsymbol{\eta}}\right\} = \begin{pmatrix} U_\mu & U_\beta \\ \mathbf{0} & S_\beta \end{pmatrix}$, and

$$U_\mu = E(\partial U / \partial \mu) = -\frac{1}{N} \sum_{i \in FP} \pi_i^{(c)} \widetilde{w}_i = -1,$$

$$U_\beta = E(\partial U / \partial \boldsymbol{\beta}^T) = \frac{1}{N} \sum_{i \in FP} \pi_i^{(c)} (y_i - \mu) \frac{\partial \widetilde{w}_i}{\partial \boldsymbol{\beta}^T} = -\frac{1}{N} \sum_{i \in FP} (y_i - \mu) x_i^T$$

$$S_\beta = E(\partial \tilde{S}^*/\partial \beta) = -\frac{1}{N+n_c}\sum_{i\in FP}\pi_i^{(c)}\cdot p_i(1-p_i)x_i x_i^T - \frac{1}{N+n_c}\sum_{i\in FP}p_i(1-p_i)x_i x_i^T$$

$$= -\frac{1}{N+n_c}\sum_{i\in FP} p_i x_i x_i^T \text{ (negative definite by condition } \mathbf{C4}\text{)}$$

It follows that $\hat{\mu} = \mu_0 + O_p(n_c^{-1/2})$, and

$$Var(\hat{\eta}) \doteq [E\{\phi(\boldsymbol{\eta}_0)\}]^{-1} Var\{\Phi(\boldsymbol{\eta}_0)\}[E\{\phi(\boldsymbol{\eta}_0)\}^T]^{-1}, \tag{B.3}$$

where $[E\{\phi(\boldsymbol{\eta})\}]^{-1} = \begin{pmatrix} -1 & \frac{N+n_c}{N}\boldsymbol{b}^T \\ \mathbf{0} & S_\beta^{-1} \end{pmatrix}$, and $\boldsymbol{b}^T = \{\sum_{i\in FP}(y_i - \mu)x_i^T\}\{\sum_{i\in FP} p_i x_i x_i^T\}^{-1}$. The middle part of (B.3), i.e., $Var\{\Phi(\boldsymbol{\eta}_0)\}$, can be calculated by partitioning $\Phi(\boldsymbol{\eta}) = \Phi_1 + \Phi_2$, where

$$\Phi_1 = \sum_{i\in FP}\begin{Bmatrix}\frac{1}{N}\delta_i^{(c)}\tilde{w}_i(y_i-\mu) \\ \frac{1}{N+n_c}\delta_i^{(c)}(1-p_i)x_i\end{Bmatrix}, \Phi_2 = \frac{-1}{N+n_c}\sum_{i\in FP}\begin{Bmatrix}0 \\ \delta_i^{(p)}d_i p_i x_i\end{Bmatrix}.$$

Notice that $\Phi_1$ and $\Phi_2$ are independent under condition **C5**, because $\Phi_1$ only involves randomization of cohort participation while $\Phi_1$ only involves survey sample selection. Hence, $Var\{\Phi(\boldsymbol{\eta}_0)\} = Var(\Phi_1) + Var(\Phi_2)$ where

$$Var(\Phi_1) = \sum_{i\in FP} p_i(1-2p_i)\begin{Bmatrix}\frac{1}{N^2}(y_i-\mu)^2/p_i^2 & \frac{1}{N(N+n_c)}(y_i-\mu)x_i^T/p_i \\ \frac{1}{N(N+n_c)}(y_i-\mu)x_i/p_i & \frac{1}{(N+n_c)^2}x_i x_i^T\end{Bmatrix}$$

under the assumption of Poisson sampling of the nonprobability sample, and

$$Var(\Phi_2) = \begin{pmatrix}0 & \mathbf{0}^T \\ \mathbf{0} & \boldsymbol{D}\end{pmatrix},$$

with $\boldsymbol{D}$ being the design-based variance-covariance matrix under the probability sampling design for sample $s_s$. For example, if survey sample is randomly selected by Poisson sampling, $\boldsymbol{D} = (N+n_c)^{-2}\sum_{i\in FP}(d_i-1)p_i^2 x_i x_i^T$.

The finite population variance of $\hat{\mu}^{ALP}$ is the first diagonal element of $Var(\hat{\eta})$, and given by

$$Var(\hat{\mu}^{ALP}) = \begin{pmatrix} -1 & \boldsymbol{b}^T \end{pmatrix} \cdot \left( Var(\Phi_1) + Var(\Phi_2) \right) \cdot \begin{pmatrix} -1 \\ \boldsymbol{b} \end{pmatrix}$$

$$= N^{-2} \sum_{i \in FP} p_i(1 - 2p_i) \left\{ \frac{(y_i - \mu)}{p_i} - \boldsymbol{b}^T \boldsymbol{x}_i \right\}^2 + \boldsymbol{b}^T \boldsymbol{D} \boldsymbol{b}.$$

Note $p_i = P(i \in s_c^* | s_c^* \cup FP) \leq 1/2$.

## C. Comparing Orders of Magnitude of $Var(\hat{\mu}^{ALP})$ and $Var(\hat{\mu}^{CLW})$

The pseudo-weighted nonprobability sample estimator of the population mean is written as

$$\hat{\mu} = \frac{1}{\sum_{i \in s_c} \widetilde{w}_i} \sum_{i \in s_c} \widetilde{w}_i y_i$$

where $\widetilde{w}_i$ is the pseudoweight $\widetilde{w}_i^{ALP}$ in the ALP estimator $\hat{\mu}^{ALP}$

$$\widetilde{w}_i^{ALP} = \frac{1 - \hat{p}_i}{\hat{p}_i} = \exp^{-1}(\hat{\boldsymbol{\beta}}^T \boldsymbol{x}_i)$$

or the pseudoweight $\widetilde{w}_i^{CLW}$ in the CLW estimator $\hat{\mu}^{CLW}$

$$\widetilde{w}_i^{CLW} = \frac{1}{\hat{\pi}_i^{(c)}} = 1 + \exp^{-1}(\hat{\boldsymbol{\gamma}}^T \boldsymbol{x}_i)$$

where $\hat{\boldsymbol{\beta}}$ and $\hat{\boldsymbol{\gamma}}$ are solutions of pseudo estimation equations $\tilde{S}^*(\boldsymbol{\beta}) = 0$ and $\tilde{S}(\boldsymbol{\gamma}) = 0$ in formulae (2.3.7) and (2.2.7) in the main text, respectively.

According to the law of total variance, finite population variance of $\hat{\mu}$ can be written as

$$V(\hat{\mu}) = E_w[V_c(\hat{\mu}|\widetilde{\boldsymbol{w}})] + V_w[E_c(\hat{\mu}|\widetilde{\boldsymbol{w}})] \tag{C.1}$$

where $\widetilde{\boldsymbol{w}} = (\widetilde{w}_1, \ldots, \widetilde{w}_N)$ is the vector of pseudo nonprobability sample weight for the finite population; $E_w$ and $V_w$ are with respect to the propensity model; $V_c$ and $E_c$ are with respect to the nonprobability sampling process, and we have

$$E_c(\hat{\mu}|\widetilde{\boldsymbol{w}}) = \frac{\sum_{i \in FP} \pi_i^{(c)} \widetilde{w}_i y_i}{\sum_{i \in FP} \pi_i^{(c)} \widetilde{w}_i} + O(n_c^{-1}) \text{ and}$$

$$V_c(\hat{\mu}|\widetilde{\boldsymbol{w}}) = \frac{\sum_{i \in FP} \pi_i^{(c)} \left(1 - \pi_i^{(c)}\right) \widetilde{w}_i^2 \left( y_i - \frac{\sum_{i \in FP} \pi_i^{(c)} \widetilde{w}_i y_i}{\sum_{i \in FP} \pi_i^{(c)} \widetilde{w}_i} \right)^2}{\left( \sum_{i \in FP} \pi_i^{(c)} \widetilde{w}_i \right)^2}$$

assuming Poisson sampling. The first term in (C.1), which is $E_w[V_c(\hat{\mu}|\widetilde{w})]$, has order $O(n_c^{-1})$ for both $\hat{\mu}^{ALP}$ and $\hat{\mu}^{CLW}$ under condition **C2**. The second term in (C.1) is approximately

$$V_w[E_c(\hat{\mu}|\widetilde{w})] \doteq \left(\frac{\partial E_c(\hat{\mu}|\widetilde{w})}{\partial \widetilde{w}}\right) V(\widetilde{w}) \left(\frac{\partial E_c(\hat{\mu}|\widetilde{w})}{\partial \widetilde{w}}\right)^T \quad (C.2)$$

The middle term in (C.2) is

$$V(\widetilde{w}) = \left(\frac{\partial \widetilde{w}}{\partial \widehat{\mathbf{B}}}\right) V(\widehat{\mathbf{B}}) \left(\frac{\partial \widetilde{w}}{\partial \widehat{\mathbf{B}}}\right)^T = \left\{\frac{\partial}{\partial \widehat{\mathbf{B}}} \exp^{-1}(\widehat{\mathbf{B}}^T x)\right\} \{V(\widehat{\mathbf{B}})\} \left\{\frac{\partial}{\partial \widehat{\mathbf{B}}} \exp^{-1}(\widehat{\mathbf{B}}^T x)\right\}^T.$$

where $\widehat{\mathbf{B}} = \widehat{\boldsymbol{\beta}}$ or $\widehat{\boldsymbol{\gamma}}$ are solutions of pseudo estimating equations $\widetilde{S}^*(\boldsymbol{\beta}) = 0$ and $\widetilde{S}(\boldsymbol{\gamma}) = 0$ in the formulae (2.3.7) and (2.2.7). Therefore

$$V_w[E_c(\hat{\mu}|\widetilde{w})] \doteq \left(\frac{\partial E_c(\hat{\mu}|\widetilde{w})}{\partial \widetilde{w}} \frac{\partial \widetilde{w}}{\partial \widehat{\mathbf{B}}}\right) V(\widehat{\mathbf{B}}) \left(\frac{\partial E_c(\hat{\mu}|\widetilde{w})}{\partial \widetilde{w}} \frac{\partial \widetilde{w}}{\partial \widehat{\mathbf{B}}}\right)^T$$

where

$$\frac{\partial E_c(\hat{\mu}|\widetilde{w})}{\partial \widetilde{w}} = \left\{\pi_1^{(c)} \frac{y_i - E_c(\hat{\mu}|\widetilde{w})}{\sum_{i \in FP} \pi_1^{(c)} \widetilde{w}_1}, \cdots, \pi_N^{(c)} \frac{y_i - E_c(\hat{\mu}|\widetilde{w})}{\sum_{i \in FP} \pi_N^{(c)} \widetilde{w}_N}\right\}^T,$$

and

$$\left(\frac{\partial E_c(\hat{\mu}|\widetilde{w})}{\partial \widetilde{w}} \frac{\partial \widetilde{w}}{\partial \widehat{\mathbf{B}}}\right) = -\frac{\sum_{i \in FP}\left\{\pi_i^{(c)} \exp^{-1}(\widehat{\mathbf{B}}^T x_i)(y_i - E_c(\hat{\mu}|\widetilde{w})) x_i\right\}}{\sum_{i \in FP} \pi_i^{(c)} \widetilde{w}_i} = O(1)$$

for both ALP and CLW.

To solve the order of $V(\widehat{\mathbf{B}})$, we first write

$$\widehat{\mathbf{B}} - \mathbf{B} = I^{-1}(\mathbf{B}) S(\widehat{\mathbf{B}}) + o_p\left(S(\widehat{\mathbf{B}})\right), \quad (C.3)$$

where $\boldsymbol{B} = \boldsymbol{\beta}$ or $\boldsymbol{\gamma}$ are solutions to the census estimating equation $S(\mathbf{B}) = 0$, and $I(\mathbf{B}) = \frac{\partial S}{\partial \mathbf{B}}(\mathbf{B})$ is the Hessian matrix.

Specifically, for the ALP method the census estimating equation can be obtained by rewriting expression (2.3.7) in the main text and differentiating with respect to $\boldsymbol{\beta}$, leading to

$$S(\boldsymbol{\beta}) = \frac{1}{N + n_c} \sum_{i \in s_c^* \cup FP} \{R_i - p_i(\boldsymbol{\beta})\} x_i,$$

where $R_i$ indicates the membership of $s_c^*$ in $s_c^* \cup FP$ (=1 if $i \in s_c^*$; 0 if $i \in FP$), and $p_i(\boldsymbol{\beta}) = E(R_i \mid x_i; \boldsymbol{\beta}) = \text{expit}(\boldsymbol{\beta}^T x_i)$ defined in Section 2.3 in the main text respectively.

The estimate $\widehat{\boldsymbol{\beta}}$ is solution to the pseudo estimating equation $\tilde{S}^*(\boldsymbol{\beta}) = 0$, where $d_i$ is the basic design weights for $i \in s_p$ and $d_i = 1$ for $i \in s_c$. We have

$$\tilde{S}^*(\widehat{\boldsymbol{\beta}}) = \frac{1}{N+n_c} \sum_{i \in s_c \cup^* s_p} d_i \{R_i - p_i(\widehat{\boldsymbol{\beta}})\} x_i = S(\widehat{\boldsymbol{\beta}}) + O_p\left(\frac{1}{\sqrt{n_c + n_p}}\right) = 0,$$

under condition **C3**, where the union $\cup^*$ allows for duplicated units in $s_c$ and $s_p$. Combined with (C.3), this leads to $\widehat{\boldsymbol{\beta}} - \boldsymbol{\beta} = O_p\left(\frac{1}{\sqrt{n_c+n_p}}\right)$ with

$$I(\boldsymbol{\beta}) = \frac{\partial S}{\partial \boldsymbol{\beta}}(\boldsymbol{\beta}) = -\frac{1}{N+n_c} \sum_{i \in s_c^* \cup FP} p_i(\boldsymbol{\beta})\{1 - p_i(\boldsymbol{\beta})\} x_i = O(1)$$

under Condition **C4**. We have

$$V(\widehat{\boldsymbol{\beta}}) = O\left(\frac{1}{n_c + n_p}\right).$$

For the CLW method, the census estimating equation is

$$S(\boldsymbol{\gamma}) = \frac{1}{N} \sum_{i \in FP} \{\delta_i - \pi_i^{(c)}(\boldsymbol{\gamma})\} x_i$$

where $\delta_i$ is the indicator of the population unit $i$ being included in $s_c$ (=1 if $i \in s_c$; 0 otherwise), and $\pi_i(\boldsymbol{\gamma}) = E(\delta_i \mid x_i; \boldsymbol{\gamma}) = \text{expit}(\boldsymbol{\gamma}^T x_i)$.

The estimate $\widehat{\boldsymbol{\gamma}}$ is solution to the pseudo estimating equation $\tilde{S}(\boldsymbol{\gamma}) = 0$ shown below

$$\tilde{S}(\widehat{\boldsymbol{\gamma}}) = \frac{1}{N}\left\{\sum_{i \in s_c} x_i - \sum_{i \in s_p} d_i \pi_i^{(c)}(\widehat{\boldsymbol{\gamma}}) x_i\right\} \quad (\text{C.4})$$

$$= \frac{1}{N}\sum_{i \in FP} \delta_i x_i + \frac{1}{N}\sum_{i \in s_p} d_i \{\delta_i^{(c)} - \widehat{\pi}_i^{(c)}\} x_i - \frac{1}{N}\sum_{i \in s_p} d_i \delta_i^{(c)} x_i = 0.$$

Under condition **C3**, we have the second and third term in (C.4)

$$\frac{1}{N}\sum_{i \in s_p} d_i\left(\delta_i - \widehat{\pi}_i^{(c)}\right) x_i = \frac{1}{N}\sum_{i \in FP}\left(\delta_i - \widehat{\pi}_i^{(c)}\right) x_i + O_p(n_p^{-1/2}), \text{ and}$$

$$\frac{1}{N}\sum_{i \in s_p} d_i \delta_i x_i = \frac{1}{N}\sum_{i \in FP} \delta_i x_i + O_p(n_p^{-1/2}).$$

Hence

$$\tilde{S}(\widehat{\boldsymbol{\gamma}}) = S(\widehat{\boldsymbol{\gamma}}) + O_p(n_p^{-1/2}) = 0,$$

which, combined with (C.3), leads to $\widehat{\boldsymbol{\gamma}} - \boldsymbol{\gamma} = O_p(n_p^{-1/2})$ with

$$I(\gamma) = -\frac{1}{N}\sum_{i \in FP} \pi_i^{(c)}(\gamma)\{1 - \pi_i^{(c)}(\gamma)\}x_i^T x_i = O(1)$$

under condition **C6** in Chen, Li & Wu (2019).

We have

$$V(\hat{\gamma}) = O\left(\frac{1}{n_p}\right)$$

As the result, the second term in (C.1) for the ALP and the CLW method has the order of $O\left(\frac{1}{n_p+n_c}\right)$ and $O\left(\frac{1}{n_p}\right)$, respectively. Combining the two terms in (C.1), we have

$$V(\hat{\mu}^{ALP}) = O\left(\frac{1}{n_p}\right) + O\left(\frac{1}{n_p+n_c}\right) = O\left(\frac{1}{n_c}\right)$$

and

$$V(\hat{\mu}^{CLW}) = O\left(\frac{1}{n_c}\right) + O\left(\frac{1}{n_p}\right) = O\left(\frac{1}{\min(n_c, n_p)}\right).$$

Therefore, in large samples we have $V(\hat{\mu}^{ALP}) \leq V(\hat{\mu}^{CLW})$, and the estimator $\hat{\mu}^{ALP}$ is more efficient than $\hat{\mu}^{CLW}$ especially when $n_c \gg n_p$.

### D. Supplementary table on estimated coefficients of propensity models

|  | RDW | CLW | ALP (FDW) | ALP.S |
|---|---|---|---|---|
| **(Intercept)** | -8.92 | -8.92 | -8.92 | 0.05 |
| **Age** (in years) | -0.06 | -0.06 | -0.06 | -0.06 |
| **Age²** | 0.00 | 0.00 | 0.00 | 0.00 |
| **Sex** (ref: male) | | | | |
| Female | -0.10 | -0.10 | -0.10 | -0.03 |
| **Education level** | -0.16 | -0.16 | -0.16 | -0.11 |
| **Race/Ethnicity** (ref: NH-White) | | | | |
| NH-Black | 1.33 | 1.33 | 1.33 | 1.47 |
| Hispanic | 1.62 | 1.62 | 1.62 | 1.64 |
| NH-Other | -0.35 | -0.35 | -0.35 | -0.28 |
| **Poverty** (ref: No) | | | | |
| Yes | 0.15 | 0.15 | 0.15 | 0.11 |
| Unknown | -0.01 | -0.01 | -0.01 | 0.01 |
| **Health Status** | 0.24 | 0.24 | 0.24 | 0.24 |
| **Region** (ref: Northeast) | | | | |
| Midwest | 0.25 | 0.25 | 0.25 | 0.15 |
| South | 0.41 | 0.41 | 0.41 | 0.35 |
| West | 0.29 | 0.29 | 0.29 | 0.14 |
| **Marital Status** (ref: married or living as married) | | | | |
| Single | -0.19 | -0.19 | -0.19 | -0.12 |
| Previously married | -0.01 | -0.01 | -0.01 | -0.02 |
| **Smoking** (ref: Non-smoker) | | | | |
| Former smoker | 0.12 | 0.12 | 0.12 | 0.10 |
| Current smoker | 0.16 | 0.16 | 0.16 | 0.14 |
| **Household Income** | -0.01 | -0.01 | -0.01 | -0.01 |
| **Chewing tobacco** (ref: No) | | | | |
| Yes | -0.35 | -0.35 | -0.35 | -0.34 |
| **BMI** (ref: normal) | | | | |
| Under-weight | -0.02 | -0.02 | -0.02 | -0.12 |
| Over-weight | 0.03 | 0.03 | 0.03 | 0.01 |
| Obese | -0.06 | -0.06 | -0.06 | -0.04 |

**Reference**

Chen, Y., Li, P., Wu, C. (2019) Doubly Robust Inference With Nonprobability Survey Samples. Journal of the American Statistical Association.; 1-11.

Krewski, D., Rao, J.N. (1981) Inference from stratified samples: properties of the linearization, jackknife and balanced repeated replication methods. The Annals of Statistics, 1010-9.

Tsiatis, A. (2007). Semiparametric theory and missing data. Springer Science & Business Media.

Wang, L. (2020) Improving external validity of epidemiologic analyses by incorporating data from population-based surveys. Doctoral dissertation, University of Maryland, College Park.